
------------------------------------------------------------------------------
\overfullrule=0.0pt
\magnification=1200

\font\eightrm=cmr8
\font\tit=cmbx10 scaled\magstep 2

\font\aun=cmbx10
\font\rauth=cmr10

\font\eightmit=cmmi8
\scriptfont4=\seveni
\vsize=20.0truecm
\hsize=14.0truecm
\hoffset=1.0 truecm
\voffset=0.0 truecm
\baselineskip=16pt plus 2pt minus 1pt
\parskip=2pt minus 1pt
\parindent=1.0truecm
\tolerance=1000

\def\a {\alpha}

\def\m {\mu}

\def\G {\Gamma}

\def\r {\rho}

\def\l {\lambda}

\def\ti {\widetilde}
\def\ltord{\hbox{$\;\raise.4ex\hbox{$<$}\kern-.75em\lower.7ex\hbox{$\sim$}
		       \;$}}
\def\gtord{\hbox{$\;\raise.4ex\hbox{$>$}\kern-.75em\lower.7ex\hbox{$\sim$}
		       \;$}}
\def\subh{\hbox{\lower.4ex\hbox{$_h$}}}
\def\subq{\hbox{\lower.2ex\hbox{$_q$}}}
\def\subR{\hbox{\lower.4ex\hbox{$_R$}}}
\def\subrhat{\hbox{\lower.2ex\hbox{$_{{\bf \hat {\it r}}}$}}}
\def\sub0hat{\hbox{\lower.2ex\hbox{$_{{\bf \hat {\it 0}}}$}}}

\def\references{\bigskip\medskip\goodbreak{\bf\noindent
References}\nobreak
\medskip\nobreak
\frenchspacing\pretolerance=2000\parindent=25truept}
\def\paper#1#2#3#4#5#6{\item{\hbox to 20truept{[#1]\hfill}}
{\rauth #2}, {#3} {\bf #4}, {#5} ({#6}). \smallskip}
\def\inpress#1#2#3#4#5{\item{\hbox to 20truept{[#1]\hfill}}
{\rauth #2}, {#3}, {#4} ({#5}). \smallskip}
\def\preprint#1#2#3#4{\item{\hbox to 20truept{[#1]\hfill}}
{\rauth #2} {#3} {\it #4} \smallskip}
\def\book#1#2#3#4#5{\item{\hbox to 20truept{[#1]\hfill}} {\rauth #2}
{#3} {\it #4}, #5 \smallskip}
\def\centra#1{\vbox{\rightskip=0pt plus1fill\leftskip=0pt plus1fill #1}}
\def\title#1{\baselineskip=30truept\parindent=0pt\centra{\tit #1}
\bigskip\baselineskip=16pt\def\titolo{#1}}
\def\first#1#2{{\aun #1$^{#2}$}}
\def\last#1#2{ \& {\aun #1$^{#2}$}}
\def\authors#1{\bigskip\centra{#1}}
\long\def\addresses#1{\bigskip\bigskip\centra{#1}}
\long\def\addr#1#2{$^{#1}${#2}\par}
\null
{\nopagenumbers
\vskip 1.0 truecm

\title{Hydrodynamics of the cosmological quark-hadron transition in
the presence of long-range energy and momentum transfer}
\authors{\first{John C. Miller}{1,2,3}\last{Luciano Rezzolla}{1}}
\addresses{
\addr{1}{Scuola Internazionale Superiore di Studi Avanzati, Trieste, Italy.}
\addr{2}{Department of Physics, University of Oxford, England.}
\addr{3}{Osservatorio Astronomico di Trieste, Trieste, Italy.}}
\vskip 1.0 truecm

\centerline {{\bf Abstract}}
\bigskip
\noindent
Two previous papers in this series have presented a study of the growth
of hadronic bubbles during the cosmological Quark--Hadron transition,
treating the material within each phase as a single perfect fluid.
Here, we extend the analysis to include the effects of long-range
energy and momentum transfer by weakly and electromagnetically
interacting particles. After a short review of the formalism adopted,
we discuss the numerical strategies used in the computer code which has
been constructed in order to solve this system of equations. Results
for the growth of single hadronic bubbles are also presented.
\vskip 0.5 truecm
\noindent
PACS Nos.: 47.55.Dz, 47.75.+f, 95.30.Jx, 98.80.Cq
\vskip 1.5 truecm
\centerline { SISSA Ref. 161/94/A-EP (Oct 94)}
\vfill\eject
}

\vsize=24.0truecm
\hsize=15.0truecm
\hoffset=0.5 truecm
\voffset=0.0 truecm
\parindent=1.0truecm
\pageno=1

\bigskip
\bigskip
\noindent
{\bf  I. Introduction }
\bigskip

	At about 10 $\m$s after the big bang it is thought that strongly
interacting matter underwent a transition from a plasma of free quarks
and gluons to one in which the quarks were confined within hadrons
(predominantly pions). Current lattice-gauge calculations [1] favour the
view that this change may well have been a continuous one, but the
possibility of it being a first order phase transition is by no means
ruled out and considerable interest continues to be focussed on the
astrophysical consequences which might have arisen if it was, indeed,
first order (see, for example [2--4]).

	The work described here (which is part of an ongoing research programme
[5--8]) is within the scenario of a first-order transition starting with the
nucleation of hadronic bubbles in a slightly supercooled quark-gluon plasma.
The bubbles then proceed to grow (with the quark-gluon plasma being
progressively transformed into a hadronic one) until eventually they coalesce
and give rise to disconnected quark regions which then shrink and probably
disappear completely (possibly leaving behind a significant baryon-number
inhomogeneity) [9] but might instead reach a stable configuration composed of
{\it strange quark matter} [10].

	In the earlier papers of the series, attention has been focussed
on studying the hydrodynamics of the growth of single hadronic bubbles
during the initial stages of the transition where it makes sense to
consider the material in each phase as a perfect fluid composed only of
the strongly interacting matter. While the transition involves only these
particles in a direct way, an important role is also played by other
particles present, which can interact with them through the
electromagnetic and weak interactions: primarily photons, electrons, muons
and their antiparticles (electromagnetic and weak interactions) and
neutrinos and antineutrinos (weak interaction only). All of these have
mean free paths long compared with that of the strongly interacting matter
and can provide a mechanism for long-range transport of energy and
momentum through the strongly interacting fluid. Since the leptons
concerned are essentially massless, both they and the photons can be
treated as components of a generalized ``radiation fluid'' and the problem
is then one of relativistic radiative transfer.

	During bubble growth, the effect of this transport becomes
significant when the radius $R_s$ of the bubble surface becomes roughly
comparable with the mean free path $\l$ of the particles concerned ($\sim
10^4$ fermi for the electromagnetic interaction and $\sim 1$ cm for the
weak interaction). When $R_s \ll \l$, the bubble is essentially
transparent to the radiation, which can then be ignored, while when $R_s
\gg \l$ the coupling is essentially complete on relevant length-scales so
that the radiation and strongly interacting matter move together as a
single fluid. Clearly, the process of coupling can, in principle, occur
twice during the bubble growth, but because the behaviour is similar in
each case, we will discuss here only the one occurring between the
strongly interacting and electromagnetically interacting particles.
Identical considerations apply also for the coupling with the neutrinos,
the only difference, apart from the scale, being the different number of
degrees of freedom into which the energy liberated by the transition is
channeled.\par

	This paper is concerned with adding the effects of radiative
transfer to our earlier scheme of calculation for following bubble growth
[7], thus extending its range of validity. Subsequently, we aim to apply a
similar scheme to the last stage of the transition (shrinking away of the
disconnected quark regions) which is likely to give rise to the most
interesting consequences from an astrophysical point of view.  The
formalism used for handling the radiative transfer has been the subject of
an earlier paper [8] and only a brief outline of this will be presented
again here.

	The plan of the paper is as follows. In Section II we give a short
review of the equations and then Section III describes the outline of our
computational scheme which uses a combined characteristic and Lagrangian
finite-difference approach. Some particular problems arise in the
computational implementation and we explain how these have been overcome
by means of introducing new variables for the radiation fluid. Section IV
concerns the setting up of initial conditions, Section V describes tests
carried out and results obtained and Section VI is the conclusion.
Throughout, we use a system of units in which $c = \hbar = k_{_B} = 1$ and
a space-like signature $(-,+,+,+)$. Greek indices are taken to run from 0
to 3 and Latin indices from 1 to 3; covariant derivatives are denoted with
a semi-colon and partial derivatives with a comma.

\bigskip
\bigskip
\noindent
{\bf  II. Relativistic Hydrodynamical Equations}
\bigskip

This section contains a short review of the equations used in the rest
of the paper. A full discussion of these has already been presented in
[8] to which the reader is referred for further details and references.

The transfer of energy and momentum is considered as taking place between
a {\it standard fluid}, composed of the strongly interacting particles and
any other particles effectively moving together with them, and a {\it
radiation fluid} consisting of those particles responsible for the
long-range transfer of energy and momentum. We treat the transfer using
the PSTF (Projected Symmetric Trace Free) tensor formalism of Thorne [11],
[12] in which the relativistic radiative transfer equation is rewritten in
terms of an infinite hierarchy of differential equations involving PSTF
moments which is then truncated by introducing suitable closure relations.
For spherical symmetry, the tensor formalism becomes effectively a scalar
one. We made the truncation at second order and obtained
$$
(w_0)_{,t}+{a\over b}(w_1)_{,\mu} +{4\over 3}\biggl({{b_{,t}\over b}+
{2R_{,t}\over R}}\biggr)w_0 +{2a\over b}\biggl({{a_{,\mu}\over a}+
{R_{,\mu}\over R}}\biggr)w_1 +\biggl({{b_{,t}\over b}-
{R_{,t}\over R}}\biggr)w_2
= as_0 ,
\eqno(1)
$$
$$
(w_1)_{,t}+{a\over b}\biggl({{1\over 3}w_0+w_2}\biggr)_{\!\! ,\mu} +
{4a_{,\mu}\over {3b}}w_0
+2\biggl({{b_{,t}\over b}+{R_{,t}\over R}}\biggr)w_1
+{a\over b}\biggl({ {a_{,\mu}\over a}+{3R_{,\mu}\over R}}\biggr)w_2
 = as_1 ,
\eqno(2)
$$
$$
w_2 = f_{\!_E} w_0 ,
\eqno(3)
$$
where $w_0$ is the energy density of the radiation, $w_1$ is the
radiation flux and $w_2$ is the anisotropy scalar of the radiation (all
measured in the local rest frame of the standard fluid); $s_0$ and
$s_1$ are energy and momentum source functions and $a$ and $b$ are
metric coefficients of the spherically symmetric line element

$$
ds^2 = -a^2 dt^2 + b^2 d\mu^2 + R^2 ( d\theta^2 +
sin^2 \theta \ d\varphi^2) ,
\eqno(4)
$$
where $\mu$ is a comoving radial coordinate and $R$ is the associated
Eulerean coordinate (the Schwarzschild circumference coordinate).
\par

The quantity $f_{\!_E}$ appearing in the closure relation (3) is a {\it
variable Eddington factor} which can take values ranging from $0$, for
complete isotropy (which could be caused by the medium being extremely
optically thick), to $2/3$ for complete anisotropy (which might arise
when the medium is very optically thin). The expression used for it has
to be arrived at on the basis of physical considerations and for our
case we have used

$$
f_{\!_E} \equiv {8 u^2 / 9 \over {(1 + 4 u^2/3)}}\left({\lambda \over
{\lambda + R}}\right) ,
\eqno(5)
$$
where $\lambda$ is the effective mean-free-path of the radiation
particles. In our picture, the bubble is first nucleated within a uniform
and isotropic quark medium and the radiation field (which interacts with
the quark medium on suitably large scales and is in thermal equilibrium
with it) initially shares these properties ({\it i.e.} $w_0$ is constant
everywhere and $w_1$ and $w_2$ are zero). The radius of the newly-formed
bubble is small compared with $\lambda$ and the medium is essentially
transparent to the radiation on this scale.  As the bubble starts to
expand, the radiation quantities deviate from their values at the time of
nucleation primarily as a result of the Doppler effect arising from the
motion of the standard fluid rest frames with respect to that of the
radiation field (which is, so far, remaining uniform in its own frame).
These Doppler corrections can be calculated analytically and the results
were presented in the Appendix of [7]. Solely on the basis of this
consideration, one finds that $f_{\!_E}$ is given by \hbox{$(8/9) u^2 (1 +
4 u^2/3)^{-1}$.} As the bubble grows to dimensions comparable with
$\lambda$, there is progressive coupling between the radiation and the
standard fluid on the relevant length-scales and this interaction tends in
the direction of making the radiation more isotropic as seen from the
standard fluid. This effect is approximated by multiplying the Doppler
term in (5) by a correction factor (in the large brackets) which has the
effect of producing the right behaviour in the optically thin and
optically thick limits and giving a physically plausible join between
them.

	The appropriate value of $\lambda$ is not known accurately but, on
the basis of elementary considerations, we have taken $\lambda \approx
10^4 \ {\rm fm}$. Tests made in order to investigate the sensitivity of
the numerical code to the values adopted for the various parameters will
be discussed in Section V.

	For the source moments $s_0$ and $s_1$ we use the expressions
$$
s_0 = {1 \over {\lambda}} (\epsilon - w_0) +
\left( s_0 \right)_{_{SC}},
\eqno(6)
$$
$$
s_1 = - {w_1 \over {\lambda}} ,
\eqno(7)
$$
where $(s_0)_{_{SC}}$ is a term expressing the contribution due to
scatterings and $\epsilon$ is the energy density for radiation in
thermal equilibrium with the standard fluid. Assuming that it roughly
follows a \hbox{black-body} law, $\epsilon$ can be written as
$$
\epsilon = g_{_R} \left({\pi^2 \over {30}}\right) T_{_F}^4 ,
\eqno(8)
$$
with $g_{_R}$ being the number of degrees of freedom of the radiation
fluid and $T_{_F}$ the temperature of the standard fluid.
\par

Obtaining a suitable expression for $(s_0)_{_{SC}}$ is less
straightforward.  While detailed derivations have been made for simpler
applications [13--14], the lack of precise knowledge about the
interaction processes in the present case, has led us express
$(s_0)_{_{SC}}$ by the simple absorption and emission term
$$
(s_0)_{_{SC}} ={\a_2
\over {\lambda}} (\epsilon - w_0),
\eqno(9)
$$
where $\a_2$ is an adjustable coefficient ranging between zero and one.
Fortunately, the results of the numerical calculations turn out not to
depend sensitively on the value chosen; a discussion of this will be
given in Section V. \par

Equations (1)--(3), describing the processes of transfer between
the standard fluid and the radiation fluid, need to be solved together
with the hydrodynamic equations
$$
u_{,t}=-a\biggl[{{\Gamma\over b}\biggl({p_{,\mu}+bs_1\over
{e+p}}\biggr) + 4\pi G R \biggl({p+{1\over 3}w_0 + w_2}\biggr) +
{G M\over {R^2}}}\biggr] ,
\eqno(10)
$$
$$
{(\rho R^2)_{,t}\over {\rho R^2}}=
-a\biggl({{u_{,\mu}-4\pi G b R w_1\over {R_{,\mu}}}}\biggr) ,
\eqno(11)
$$
$$
e_{,t}=w\rho_{,t}-as_0 ,
\eqno(12)
$$
$$
{(aw)_{,\mu}\over {aw}}=-{w\rho_{,\mu}-e_{,\mu}+bs_1\over {\rho w}} ,
\eqno(13)
$$
$$
M_{,\mu}=4\pi R^2 R_{,\mu}\biggl({e + w_0 +{u\over {\Gamma}}w_1}\biggr) ,
\eqno(14)
$$
$$
\Gamma=\biggl({1+u^2-{2GM\over R}}\biggr)^{\!1/2}={1\over b}R_{,\mu} ,
\eqno(15)
$$
$$
b = {1 \over { 4 \pi R^2 \rho}} .
\eqno(16)
$$
where $\rho$ is the {\it relative compression factor} (which plays the
same role as played by the rest-mass density in a non-relativistic
fluid), $\Gamma$ is the general relativistic analogue of the Lorentz
factor, and $w$ is the specific enthalpy ($w = (e+p)/ \rho$). The
generalized mass function $M$ can also be calculated using the
alternative equation
$$
M_{,t}=-4\pi R^2 R_{,t}
\biggl({p + {1\over 3}w_0 +{\Gamma\over u}w_1+w_2}\biggr) .
\eqno(17)
$$
Equations of state are required for both phases of the strongly
interacting matter. The hadronic medium is taken to consist of
massless, pointlike pions for which
$$
e\subh = g\subh \left({\pi^2 \over {30}}\right) T^4_h \hskip 4.0truecm
p\subh = {1 \over 3} e\subh ,
\eqno(18)
$$
while the quark phase is described by the M.I.T. {\it Bag Model}
equation of state [15]
$$
e\subq = g\subq
\left({\pi^2 \over {30}}\right) T^4_q + B
\hskip 4.0truecm
p\subq = g\subq \left({\pi^2 \over {90}}\right) T^4_q - B ,
\eqno(19)
$$
where $B$ is the ``Bag'' constant. We take $g\subq=37$, $g\subh=3$ and
these values need to be incremented by the relevant number of
additional degrees of freedom $g_{_R}$ when any of the radiative
particles are completely coupled to the strongly interacting matter.
\par

For treating the phase interface, we again use a characteristic scheme
(as in [7]). The characteristic form of (1), (2), (10) and (12) is
$$
\displaylines{\qquad{}
du\;{\pm}\;{\Gamma\over {\rho w c_s}}dp+
a\biggl\{{\Gamma\over {\rho w}}(s_1\;{\pm}\;c_ss_0) \hfill\cr
\hfill{}+4\pi G R\biggl[{p+\biggl({{1\over 3}+
f_{\!_E}}\biggr)w_0\;{\mp}\;c_sw_1}\biggr]+
{GM\over {R^2}}\;{\pm}\;{2\Gamma uc_s\over R}\biggr\}dt=0, \qquad\cr }
$$
\rightline{(20)}
\medskip
\noindent
which are to be solved along the forward and backward characteristics of
the standard fluid $d\mu = \pm \ (a/b) c_s\,dt$ (here $c_s=(\partial p/
\partial e)^{1/2} $ is the local sound speed in the standard fluid), and

$$
\displaylines{
dw_1\;{\pm}\;\biggl({{1\over 3}+f_{\!_E}}\biggr)^{\!1/2}dw_0
 +\biggl[{\biggl({{4\over 3}+f_{\!_E}}\biggr)w_0\;{\pm}\;{2(c_s^2-1-K)
(1/3+f_{\!_E})^{1/2}\over {c_s^2-1/3-f_{\!_E}}}w_1}\biggr]
{1\over {\Gamma}}du\cr
 +\biggl[{2(f_{\!_E}-2/3-K)\over
{\rho w(c_s^2-1/3-f_{\!_E})}}\biggr]w_1dp+a\Biggl\{
\biggl({{2u\over R}-{4\pi GRw_1\over {\Gamma}}}\biggr)\times\cr
 \times \biggl[{{2[(1/3+f_{\!_E})(c_s^2-1)-K c_s^2]\over
{c_s^2-1/3-f_{\!_E}}}w_1\;{\pm}\;\biggl({{4\over 3}+f_{\!_E}}\biggr)
\biggl({{1\over 3}+f_{\!_E}}\biggr)^{\!1/2}w_0}\biggr]\cr
 +\biggl[{4\pi R\biggl({p+w_0\biggl({{1\over 3}+f_{\!_E}}\biggr)}\biggr)+
{M\over {R^2}}}\biggr]\times \cr
 \times \biggl[{\biggl({{4\over 3}+f_{\!_E}}\biggr)w_0\;{\pm}\;
{2(c_s^2-1-K)(1/3+f_{\!_E})^{1/2}\over {c_s^2-1/3-f_{\!_E}}}w_1}\biggr]
{G\over {\Gamma}} - {K u (1+4 u^2 /3) \over {\l
(1 + R/ \l)}} w_1 \cr
 -{1\over R}\biggl\{{3f_{\!_E}\biggl[{ {\pm} \biggl({{1\over
3}+f_{\!_E}}\biggr)^{\!1/2}u - \Gamma }\biggr]w_0\;-2\;\biggl[{ {\pm}
\biggl({{1\over 3} +
f_{\!_E}}\biggr)^{\!1/2} \Gamma - u}\biggr]w_1}\biggr\}\cr
+\biggl[{{2c_s^2\over {\rho w (c_s^2-1/3-f_{\!_E})}}
\biggl({f_{\!_E}-{2\over 3} - K} \biggr)
w_1\;{\mp}\;\biggl({{1\over 3}+
f_{\!_E}}\biggr)^{\!1/2}}\biggr]s_0\; \cr
  + \;\biggl[{\pm}  {  2 \over {\rho w (c_s^2-1/3-f_{\!_E})}}
\biggl({f_{\!_E}-{2\over 3} - K}\biggr)
\biggl({{1\over 3}+
f_{\!_E}}\biggr)^{\!1/2}w_1\; - \;1\biggr] s_1\Biggr\}dt=0 ,
\cr\cr}
$$
\rightline{(21)}
\eject
\medskip
\noindent
which are to be solved along the forward and backward characteristics
of the radiation fluid $d\mu = \pm \ (a/b) (1/3 +f_{\!_E})^{1/2}dt$ ,
and where for compactness we have defined

$$
K = f_{\!_E}{ \Gamma w_0 \over{ u (1 + 4u^2 /3) w_1 }} .
\eqno(22)
$$

Supplementary equations calculating $\rho$, $R$ and $M$ are solved
along the flowlines of the standard fluid (i.e. advective directions
$d\mu = 0$) and are

$$
d\rho-{1\over {c_s^2w}}dp-{as_0\over w}dt=0 ,
\eqno(23)
$$
$$
dR = a u \, dt ,
\eqno(24)
$$
$$
dM=-4\pi R^2 a u \biggl[{p+\biggl({{1\over 3} +f_{\!_E} }\biggr)w_0
+ {\Gamma\over u}w_1}\biggr]dt .
\eqno(25)
$$

The configuration of characteristic curves adjacent to the interface is
shown in Figure 1 for evolution of the system from time level $t$ to a
subsequent time level $t + \Delta t$. The dashed lines represent the
forward and backward characteristics for the radiation fluid {\bf r},
the full narrow lines are the equivalent characteristics for the
standard fluids {\bf f}, the vertical dotted line is a flow-line of the
standard fluid in the quark phase and the heavy line is the worldline
of the interface. \par

\vskip 4.0 truecm
\centerline{
\vbox{\hsize=13.0truecm \baselineskip=12pt minus 1pt \noindent\eightrm
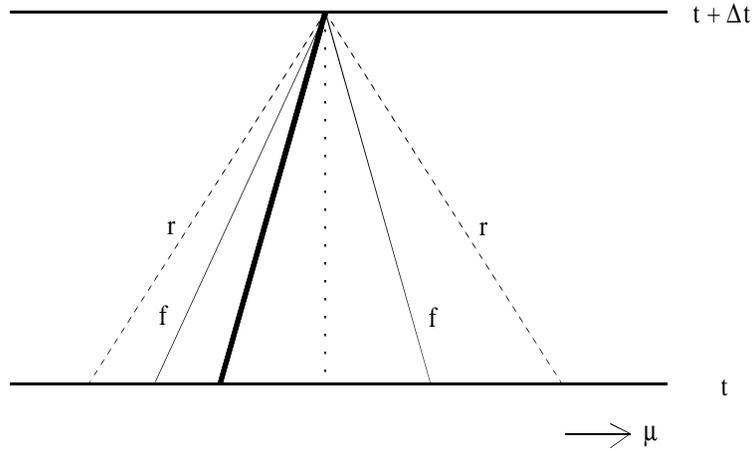
Figure 1. The configuration of characteristic curves near the phase
interface drawn in the Lagrangian coordinate frame.}}
\bigskip

Note that the difference between the characteristic directions results
from the difference between the sound speeds in the radiation fluid
($(1/3 +f_{\!_E})^{1/2}$) and in the standard fluid ($c_s$). If the
latter were not relativistic, this difference would be large but in the
present case $c_s \to 1/\sqrt {3} $ and the difference between the
sound speeds is frequently very small.  This leads to some considerable
complications in the numerical solution of the equations which we will
discuss in Section IV. \par

To complete the solution at the interface, it is also necessary to
introduce junction conditions across it and those for the energy and
momentum of the standard fluid can be obtained using the Gauss--Codazzi
formalism [16, 17].  Taking the surface tension $\sigma$ to be independent
of temperature, these junction conditions are

$$
[(e+p)ab]^{\pm}=0 ,
\eqno(26)
$$
$$
[eb^2{\dot \mu_{_S}}^2+pa^2]^{\pm}=-{\sigma f^2\over 2}\biggl\{
{{1\over {ab}}{d\over {dt}}\biggl({b^2\dot \mu_{_S}\over f}\biggr) +
{f_{,\mu}\over {ab}} +
{2\over {fR}}{(b\dot \mu_{_S} u + a\Gamma)}}\biggr\}^{\pm} ,
\eqno(27)
$$
where $[A]^{\pm} = A^+ -  A^-$,\  $\{A\}^{\pm} = A^+ + A^-$,\
$\mu_{_S}$ is the interface location, $\dot \mu_{_S} = d\mu_{_S}/dt $,
$f = (a^2 - b^2 \dot \mu^2_{_S} )^{1/2}$ and the superscripts $^{\pm}$
indicate quantities immediately ahead of and behind the interface [5].

Up to the time of the complete coupling, it is reasonable to neglect
any interaction of the radiation fluid with the matter in the phase
interface and so the energy and momentum junction conditions for the
radiation are just continuity conditions:

$$
\biggl[{ab\dot \mu_{_S}\biggl({{4\over 3}+f_{\!_E}}\biggr)
w_0-(a^2+b^2{\dot \mu_{_S}}^2)w_1}\biggr]^{\pm}=0 ,
\eqno(28)
$$
$$
\biggl[{\left\{{a^2\biggl({{1\over 3} + f_{\!_E}}\biggr)+
b^2{\dot \mu_{_S}}^2}
\right\}w_0-2ab\dot \mu_{_S} w_1}\biggr]^{\pm}=0 .
\eqno(29)
$$

Other supplementary junction conditions follow from continuity
across the interface of the metric quantities $R$, $dR/dt$, and $ds$

$$
[R]^{\pm}=0 ,
\eqno(30)
$$
$$
[au+b \dot \mu_{_S} \Gamma]^{\pm}=0 ,
\eqno(31)
$$
$$
[a^2-b^2 {\dot \mu_{_S}}^2]^{\pm}=0 ,
\eqno(32)
$$
and from the time evolution of the mass function $M$

$$
{d \over {dt}}[M]^{\pm} = 4\pi R^2_{_S} \left[ { b \Gamma {\dot \mu_{_S}}
\left\{ ( e + w_0 + {u \over \Gamma}w_1 ) \right\} - a u \left\{ { p +
\left( { {1\over 3} + f_{\!_E} } \right) w_0 + {\Gamma \over u} w_1 }
\right\} } \right]^{\pm}.
\eqno(33)
$$

At the time of bubble nucleation, conditions are essentially Newtonian
so that

$$
[M]^{\pm}=4\pi R^2_{_S} \sigma .
\eqno(34)
$$

One further equation is needed in order to complete the solution at the
interface and this is an expression for the rate at which the quark
matter passes across it.  A suitable expression is obtained by setting
the hydrodynamical flux $F_{_H}$ into the hadron region equal to the
net thermal flux $F_{_T}$ into it

$$
F_{_H}={{aw\dot\mu_{_S}}\over{4\pi R^2_{_S}(a^2-b^2\dot \mu_{_S}^2)}} =
\left({\a_1\over 4}\right) g\subh \left({\pi^2 \over {30}}\right)
(T_q^4-T_h^4)=F_{_T}
\eqno(35)
$$

\noindent
where $\alpha_1$ is an accommodation coefficient ($0\le \a_1 \le 1$). \par

This completes the set of equations. In the next Section we will
present the details of the computational scheme and discuss the
techniques and the strategies used.

\bigskip
\bigskip
\noindent
{\bf III. Numerical Computations of Bubble Growth}
\bigskip

	In order to solve the equations of the previous section, we have
constructed a computer code following the lines of the one developed
previously for calculations of the initial stages of bubble growth (see
[7] for a full description). As before, this employs a composite
numerical strategy in which a standard Lagrangian finite-difference
method is used to solve the hydrodynamical equations in the bulk of
each phase, while the system of characteristic equations and junction
conditions is solved for the grid zones immediately ahead of and behind
the phase interface (which is tracked continuously through the grid).
The radiation quantities $w_0$ and $w_2$ are taken as ``mid-zone''
quantities while the radiation flux $w_1$ is taken as a zone boundary
quantity and calculated at the half time level (as for the velocity $u$
-- see [7] for details).

	The equations presented in Section II are general in nature and can
be applied to a variety of situations. Normally, there would be no
problem in doing this but some particular difficulties have arisen when
applying them to the present case of bubble growth at the cosmological
\hbox{Q--H} phase transition. Here, direct use of the radiation
equations in the form given above leads to rapidly-growing
instabilities which destroy the solution. After a series of experiments
it was found that the difficulty originates in the very small deviation
of $w_0$ and $w_1$ away from their initial values during the early part
of the bubble growth, and in the fact that the characteristic sound
speed in the radiation fluid $(1/3+f_{\!_E})^{1/2}$ becomes very close
to the sound speed in the standard fluid $c_s$ when the radiation is
nearly isotropic in the rest frame of the standard fluid ({\it i.e.}
when $f_{\!_E} \to 0$). These features lead to production of
cancellation errors in the solution of Eqs. (1) and (2) and near
divergences in the characteristic form of the equations (21) (the
expression $(c_s^2-1/3-f_{\!_E})$ appears in the denominator of several
terms). Note that the near equality of the sound speeds only arises
when the standard fluid is, itself, relativistic (with $c_s \sim
1/\sqrt{3}$). Also, it is a peculiarity of the present situation that,
initially, the radiation is nearly isotropic in the rest frame of the
standard fluid not because the medium is optically thick on the scale
of the bubble but, rather, because of the assumed isotropy of the
universe.

	For overcoming the cancellation errors, we have introduced new
radiation variables defined as the difference between the energy density,
the flux and the shear of the radiation fluid and some reference values
(indicated below by the superscript $^*$) with the aim of performing an
analytic cancellation of large terms in the equations leaving behind
smaller ``difference'' terms. It turns out to be convenient to take these
reference values to be those which would be measured if the only effect
were that resulting from the motion of the fluid relative to a uniform
radiation field having an energy density equal to that at the time of
nucleation of the bubble, $(w_0)_{_N}$.  (These are the pure Doppler
values mentioned earlier and calculated in the Appendix of [7].) Using a
tilde to denote the new variables, we have

$$
\eqalignno{
{\ti w_0}&=w_0 - (w_0)^{*}=w_0 -
\left({1+{4\over 3} u^2 }\right)(w_0)_{_N},  &(36)\cr
{\ti w_1}&=w_1 - (w_1)^{*}=w_1+{4\over 3} u\Gamma (w_0)_{_N},&(37)\cr
{\ti w_2}&=w_2 - (w_2)^{*}=w_2 - {8\over 9}u^2(w_0)_{_N} ,&(38)\cr}
$$
\bigskip
\noindent
and equations (1) and (2) can then be rewritten as
\eject
$$
\displaylines{
\qquad{} ({\ti w_0})_{,t} +
a{\ti w_0} \left [{1 \over {R^2}}
\left ( {4\over 3}+ f_{\!_E} \right )
(uR^2)_{,R} - {3 u f_{\!_E} \over R} \right ]+
{\G \over {a R^2}} ({\ti w_1} a^2 R^2)_{,R} \hfill{} \cr
+ a {4 \over {3 R}} ({ w_0})_{_N}
\left [ f_{\!_E} ( {3\over 4}  + u^2) - {2 \over 3} u^2 \right ]
\left [{1 \over R} (u R^2)_{,R} - 3 u \right]  - a s_0 \cr
 - {4 \over 3 } a ({ w_0})_{_N} G \left [ 4 \pi u R
\left ( 2p - e - {{w_0} \over 3} + 2 {w_2} -
{u \over {\G }} {w_1} \right ) - {M \over R }
\left ( 2 u_{,R} + {u \over R}\right ) \right ]
\hfill{} \cr
\hfill{} - { 4 \pi a G R \over {\G}} \left ( {4 \over 3} {w_0} +
{w_2} \right )  w_1 =0 , \hskip 1truecm }
$$
\rightline{(39)}
$$
\displaylines{
\qquad{} ({\ti w_1})_{,t} + 2 {\ti w_1} {a \over R} (u R)_{,R} +
a \G \left ({{\ti w_0} \over 3} + {\ti w_2} \right)_{,R}+
\G \left( {4 \over 3}{\ti w_0} + {\ti w_2}\right) a_{,R}
+ {3 a \G {\ti w_2}\over R} \hfill \cr
\qquad\qquad - a s_1+
{4 \over 3 } a ({w_0})_{_N} \G G \left [ 4 \pi R
\left ( p + {{ w_0} \over 3} + { w_2} -
{u \over {\G }}{\ w_1}\right ) +
{M \over {a^2 R^2}}(a^2 R)_{,R} \right ]  \hfill \cr
\hfill - {8 \pi a G R {w_1}^2 \over {\G}}=0 , \hskip 1truecm  }
$$
\rightline{(40)}
\medskip
\noindent
where the partial derivatives with respect to $\m$ have been replaced by
the equivalent derivatives with respect to $R$ ({\it i.e.} $\partial /
\partial R = (4 \pi R^2 \rho / \G) \partial / \partial \mu $). Equations
(39) and (40) are the new radiation hydrodynamical equations for the bulk
of each phase; once the ``tilde'' variables have been computed, the values
of $w_0, w_1, w_2$ can be calculated from (36)--(38). Note that in (39),
(40) the radiation variables which are multiplied by $G$ are not
transformed according to (36)--(38). This has been done to keep the
expressions in a simpler form and because the contribution of these
terms is small under the present circumstances.

 	Using the new variables, the radiation characteristic equations
become
$$
d{\ti w_1} \pm ({1\over 3} +f_{\!_E})^{1/2} d{\ti w_0}+
{\rm BU} du + {\rm BP} dp +  {\rm BT} dt=0 ,
\eqno(41)
$$
where
$$
\displaylines {
{\rm BU}= {a\over {\G}} \Biggl\{
\left( {4\over 3} + f_{\!_E} \right) {\ti w_0}
+{8\over 3} (w_0)_{_N} \left ( {GM\over R} -
{u^2 R\over {3(\l+R)} }\right) \hfill{} \cr
 \pm \; 2 ({1\over 3}+f_{\!_E})^{1/2} \left \{
{2 \G \over u} \left [ {\ti w_0 \over {(1+4u^2/3)}} -
 ({w_0})_{_N}{R \over {\l}} \left( 1+ {4\over 3}u^2 \right) \right]
+ {4 \ti w_1 \over {3 f_{\!_E}} } \right\} \Biggr\}, \cr }
$$
\rightline{(42)}
\eject
\bigskip
$$
\displaylines {
{\rm BP}=
{2 \G \over {\r w}} \biggl[ {{\ti w_0} \over {u(1+4u^2/3)}} +
{2 {\ti w_1} \over {3 \G f_{\!_E}} } \left(1 -{3\over 2} f_{\!_E} \right)
-({w_0})_{_N}{R\over {u \l}} \left( 1+ {4\over 3}u^2 \right)
 \biggr] , \cr}
$$
\rightline{(43)}
\medskip
$$
\displaylines {
{\rm BT} = \Biggl \{ ({\rm BU})
\biggl\{ 4\pi G R\biggl[ p+\biggl( {1\over 3}+
f_{\!_E} \biggr)w_0 \biggr]+
{GM\over {R^2}} + {\Gamma \over{\rho w}} s_1 \biggr\} \cr
+ {2 a c_s^2 \over {\rho w}} ({\rm BP})
\left[ s_0+\rho w \left( {2 u\over R} -
{4\pi G R w_1\over {\G}}\right)\right] \cr
+ a \Biggl\{ {\ti w_0}
\biggl[\G f_{\!_E}{(3\l+2R)\over{R(\l+R)}}-
{s_1\over {\rho w}}\left( {4\over 3}+f_{\!_E}\right) \pm
\left({1\over 3}+f_{\!_E}\right)^{1/2} {u\over R}
\left( {8\over 3}-f_{\!_E}\right) \biggr] \hfill{} \cr
\hskip  2.0truecm + 2{\ti w_1}\left[ {u\over R}-
{4\pi G R w_1\over {\G}} \pm
\left({1\over 3}+f_{\!_E}\right)^{1/2}
\left( {\G\over R}-{s_1\over {\rho w}}\right)\right] \hfill{} \cr
\hskip 2.0truecm-{8 \G u^2 \over {9(\l+R)}}
\left[{(4\l+3R)\over{(\l+R)}} \mp
\left({1\over 3}+f_{\!_E}\right)^{1/2}
{u\over{\G}} \right]({w_0})_{_N} \hfill{} \cr
\hskip 2.0truecm +{4\over 3} ({w_0})_{_N}
\Biggl\{ 4 \pi G \G R \biggl[
\biggl( p+{w_0\over 3} +
{u\over{\G}}w_1+w_2 \biggr) \hfill{} \cr
\hskip 2.0truecm \mp {u\over{\G}} \left({1\over 3}+
f_{\!_E}\right)^{1/2}
\left(2p-e-{w_0\over 3}-
{u\over{\G}}w_1+2w_2\right)\biggr] \hfill{} \cr
\hskip 3.0truecm +{\G GM \over{R^2}}\left(1 \pm
\left({1\over 3}+f_{\!_E}\right)^{1/2}
{u\over{\G}}\right) \pm \left({1\over 3}+
f_{\!_E}\right)^{1/2} s_0\Biggr\} \hfill{} \cr
\hskip 2.0truecm -s_1 \left[ 1+ {({w_0})_{_N}\over{\rho w}}
\left({8GM\over{3R}} -
{8u^2 R\over{9(\l + R)}}\right)\right] \hfill{} \cr
\hskip 3.0truecm \mp \left({1\over 3} + f_{\!_E}\right)^{1/2}
{4\pi G R w_0w_1\over{\G}}
\left({4\over 3}+f_{\!_E}\right)
\Biggr\}\Biggr\} \cr
=
 ({\rm BU})
\biggl\{ 4\pi G R\biggl[ p+\biggl( {1\over 3}+
f_{\!_E} \biggr)w_0 \biggr]+
{GM\over {R^2}} + {\Gamma \over{\rho w}} s_1 \biggr\} \cr
+ {2 a c_s^2 \over {a \rho w}} ({\rm BP})
\left[ s_0+\rho w \left( {2 u\over R} -
{4\pi G R w_1\over {\G}}\right)\right] + a \overline {{\rm BT}}.
\hskip 1.0truecm \cr}
$$
\rightline{(44)}
\medskip
The new form of the radiation junction conditions (28) and (29) is
\eject
$$
\displaylines{
\biggl[ ab {\dot \mu_{_S}} \left( {4\over 3}+f_{\!_E} \right)
{\tilde w_0} -(a^2+b^2 {\dot \mu_{_S}}^2) {\tilde w_1}+
({w_0})_{_N} \biggl\{ a b {\dot \mu_{_S}} \left( 1 + {4 \over3 } u^2 \right)
\left( {4 \over 3} + f_{\!_E} \right) \cr
+ {4 \over 3} u \Gamma (a^2+b^2 {\dot \mu_{_S}}^2)
\biggr\}\biggr]^{\pm}=0 , \cr}
$$
\rightline{(45)}
$$
\displaylines{
\qquad \biggl[\left\{a^2\left({1\over 3} + f_{\!_E}\right)+
b^2{\dot \mu_{_S}}^2\right\}{\tilde w_0}
-2ab\dot \mu_{_S}{\tilde w_1} \hfill \cr
+ ({w_0})_{_N} \left\{\left( 1 + {4 \over3 }u^2\right)
\left[ a^2\left({1\over 3} + f_{\!_E}\right)+
b^2{\dot \mu_{_S}}^2 \right]+
{8\over 3}abu\Gamma \dot \mu_{_S} \right\} \biggr]^{\pm}=0 . \cr}
$$
\rightline{(46)}
\medskip
\noindent
Note that the characteristic equation no longer has terms with
$(c_s^2-1/3-f_{\!_E})$ in the denominator but it {\it does} have terms
containing the ratio ${\ti w_1}/f_{\!_E}$ and these still give rise to
numerical instabilities. However, this can be countered by further
rewriting the equations in a form in which ${\ti w_1}/f_{\!_E}$ only
appears as the coefficient of expressions which are small when $f_{\!_E}
\to 0$. The central point in our strategy consists in isolating the group
of terms which appears on the left-hand-side of the fluid characteristic
equations (20) (and hence tends to zero when $f_{\!_E} \to 0$ and the
fluid and radiation characteristics coincide). Details of the manipulation
involved are given in the Appendix. This group of terms can then be
conveniently handled using the differences between parameter values at the
feet of the fluid and radiation characteristics. The revised form of the
radiation characteristic equations (which is the one actually implemented
in the code) is
$$
\displaylines{
d{\ti w_1} \pm ({1\over 3} +f_{\!_E})^{1/2} d{\ti w_0}+
{\rm BU} du - {2 {\ti w_1} \over {(e+p)}} dp +
{\rm BT} dt  \hfill{} \cr
\hskip 1.0truecm + {c_{s}\over \G} \Biggl\{ {2 \G \over u}
\left [{{\ti w_0} \over {1+4u^2/3}} -
({w_0})_{_N}{R\over {\l}}
\left ( 1+ {4\over 3}u^2 \right) \right]+
{4 {\ti w_1} \over {3 f_{\!_E} }} \Biggr\} \times \hfill{} \cr
\hskip 2.0truecm\times\Biggl\{du\;{\pm}\;{\Gamma\over {\rho w c_s}}dp+
a\biggl\{{\Gamma\over {\rho w}}
(s_1\;{\pm}\;c_ss_0) \hfill\cr
\hfill{}+4\pi G R\biggl[{p+\biggl({{1\over 3}+
f_{\!_E}}\biggr)w_0\;{\mp}\;c_sw_1}\biggr]+
{GM\over {R^2}}\;{\pm}\;{2\Gamma uc_s\over R}\biggr\} dt \Biggr\}
=0 ,\qquad\cr }
$$
\rightline{(47)}
\medskip
\noindent
where
$$
\displaylines{
{\rm BU} = {a \over {\G}} \Biggl\{\left [ \left ( {4\over 3} +
f_{\!_E} \right) {\ti w_0} +
{8\over 3} ({w_0})_{_N} \left ( {GM\over R} -
{u^2 R\over {3(\l+R)} }\right)\right] \hfill{} \cr
\pm \left ( {({1/ 3} + f_{\!_E})^{1/2}  \over {c_s}} -
1 \right) {c_{s}\over \G}
\biggl \{ {2 \G \over u} \biggl[ {{\ti w_0} \over {1+4u^2/3}}
-({w_0})_{_N}{R\over {\l}}
\left ( 1+ {4\over 3}u^2 \right) \biggr]+
{4 {\ti w_1} \over {3 f_{\!_E} }} \biggr\}\Biggr\} , \cr}
$$
\rightline{(48)}
$$
\displaylines{
\hskip 2.0truecm
{\rm BT}=({\rm BU})
\Biggl\{ G \biggl\{ 4\pi R \biggl[ p +
\left({1\over 3}+f_{\!_E} \right)w_0 \biggr] +
{M\over R^2} \biggr\} +
{\G\over {\rho w}} s_1 \Biggr\} \hfill{} \cr
\hskip 1.0truecm+ {2 a \over {\rho w}}c_s^2 \left[ s_0+\rho w
\left( {2 u\over R} -
{4\pi G R w_1\over {\G}}\right)\right]
\left({4 \over 3} \G u ({w_0})_{_N} -
{\ti w_1} \right) + a \overline {{\rm BT}}, \hskip 1.0truecm \cr}
$$
\rightline{(49)}
\noindent
where $\overline {{\rm BT}}$  is the same as in (44). Note that the last
two lines of (47) are the terms on the left-hand-side of the standard
fluid characteristic equations (20).

\bigskip
\bigskip
\noindent
{\bf  IV. Initial conditions}
\bigskip
Making use of the experience gained in the earlier work [7], the setting
of initial conditions has been quite straightforward. We started with a
single supercooled hadronic bubble nucleated in mechanical and thermal
equilibrium with its surroundings at a temperature $T_{_N}$ slightly below
the critical temperature for the transition $T_{_C}$. The equilibrium is an
unstable one and any perturbation (continuing expansion of the universe,
for example) will cause it to start growing. However, this growth is
extremely slow and, in practice, it is not easy to follow with our code as
numerical noise rapidly dominates. Our strategy then was to introduce a
small artificial perturbation, decreasing the fluid temperature inside the
bubble by a small amount below its equilibrium value, and analytically
tracing the effect of this on related quantities (including the velocity
field of the standard fluid). This successfully produced a suitable data
set for starting the time evolution in a smooth and consistent way (see
[7]).

	Since, at this stage, the radius of the bubble is small compared
with $\lambda$, the radiation fluid is not significantly affected by the
thermal perturbation and remains uniform and isotropic in its own frame.
The initial conditions for $w_0, w_1, w_2$ (measured in the rest frames
of the standard fluid) are then those calculated from the Doppler
formulae discussed earlier.

\bigskip
\bigskip
\noindent
{\bf  V. Tests and results }
\bigskip
As usual in numerical computations, the construction of the computer
code was followed by a series of tests to eliminate errors and verify
that the strategies used were satisfactory. One important test consisted
in turning off the source functions and checking that the computed
values of the radiation variables agreed with the analytical Doppler
expressions. This revealed the problems discussed earlier. When these had
been satisfactorily solved, the source functions were then turned on again
and complete runs of the code were carried out. As the radius of the
bubble increased (leading to increased coupling between the radiation and
the standard fluid on relevant length scales) care was required as
increasingly steep gradients of $w_0$ appeared in the vicinity of the
interface prior to complete coupling. Since structure on a scale smaller
than the grid spacing can obviously not be resolved, it is necessary to be
ready to switch on complete coupling in the equations at the appropriate
moment.  Some experimentation was required in order to do this in the best
way.  When this had been done, further tests were carried out in order to
examine the sensitivity of the results to changes in the physical
parameters and assumptions. In the following, we will first present
results for a set of ``canonical'' parameter values and then discuss the
effect of varying some of these.

	In Figs. 2 -- 5, we show results from a run with $T_{_C} = 150$
MeV, $T_{_N}/T_{_C} = 0.98$, $\sigma _0 = 1$, $\alpha_1 = \alpha _2 = 1$
and $\l =10^4$ fm. Here $\sigma_0 = \sigma / T_{_C}^3$ is a parameter
commonly used to measure the relative strength of the surface tension.
The value which we are taking for this is larger than currently
preferred ones but we give results for this case to allow direct
comparison with those of [7]. Figs. 2 and 3 show the behaviour of the
velocity and energy density of the standard fluid, at various times
during the bubble growth, while Figs. 4 and 5 show the corresponding
behaviour of the radiation energy density and flux. All of these figures
should be viewed together and it is useful, also, to make comparison
with Figs. 2 and 4 of [7] which are for the equivalent calculation
without inclusion of the radiation particles. (Note that for convenience
in drawing these figures, the values of the variables at the centre of
the bubble have been plotted at $log_{10} R {\rm (fm)} = 0$ rather than
at R = 0.)

	During the first part of the bubble expansion ($R_s \ltord 10^2$
fm), the standard fluid variables behave in an identical way to that seen
previously in the calculation with no radiative transfer: the velocity of
the interface progressively increases and a compression wave is pushed out
into the surrounding quark medium. The velocity profile in the quark phase
is approximately solenoidal ($u \propto 1/R^2$). The radiation variables
at this stage have profiles which are almost exactly the Doppler ones
produced by the motion of the fluid relative to an essentially uniform
radiation field. In the previous calculations, where radiative transfer
was not included, the standard fluid variables tended towards a similarity
solution which was effectively attained for $R_s \gtord 10^3$ fm. In the
present calculation, the coupling together of the radiation and the
standard fluid (which can be clearly seen in Fig. 4) starts to be
effective before the former similarity solution is fully reached, causing
first a distortion of the velocity profile and subsequently a decrease in
the peak velocity. (If a smaller value is used for $\sigma_0$, both bubble
nucleation and the attainment of the similarity solution occur at smaller
values of $R_s$.) As the coupling becomes more complete on the relevant
length scales (when $R_s \sim 10^4$ fm), the peak of the radiation flux
profile becomes very narrow (the main part of the flux is concentrated
exactly at the interface) and when it is no longer possible to resolve
this on the grid we switch to total coupling. This involves setting to
zero the radiation flux $w_1$, the Eddington factor $f_{\!_E}$ and the
source functions $s_0$ and $s_1$ and augmenting the number of degrees of
freedom for the standard fluid at the interface to include also those of
the coupled radiation. The behaviour of the radiation at the interface is
then included together with that of the standard fluid. Following the
total coupling, the variables tend rapidly to a new similarity solution
characterized by smaller velocities (the front now has to push a medium
having larger inertia) and a smaller temperature jump across the
interface.

\vskip 4.0 truecm
\centerline{
\vbox{\hsize=13.0truecm \baselineskip=12pt minus 1pt
\eightrm
\textfont1=\eightmit
\scriptfont0=\fiverm
\noindent
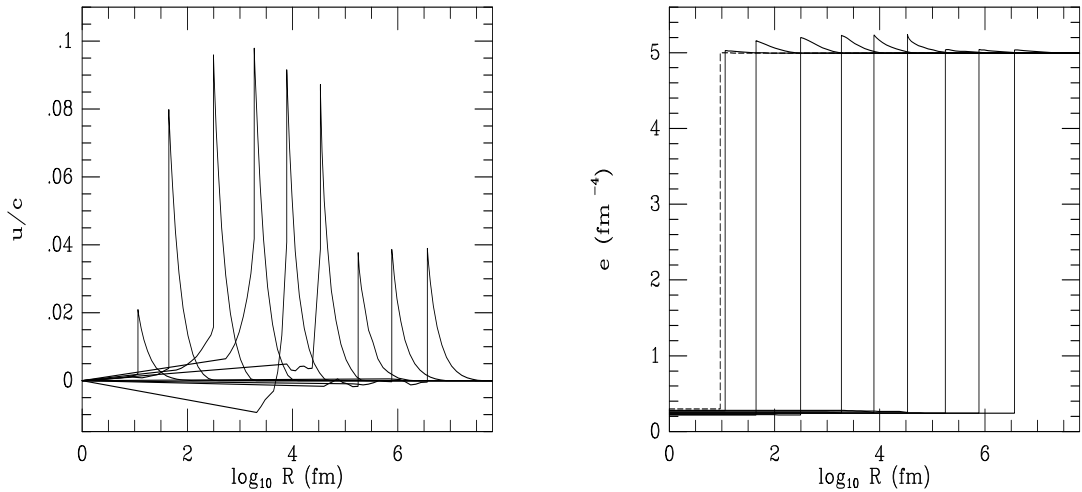
Figures 2--3. Velocity of the standard fluid $u$ and energy
density $e$. Different curves are for different times during the
bubble growth; the dashed lines represent the values at the initial
time. Here $\a_2$ = 1 and $\l$ = 10$^4$ fm. }}

\vskip 4.0 truecm
\centerline{
\vbox{\hsize=13.0truecm \baselineskip=12pt minus 1pt
\eightrm
\textfont1=\eightmit
\scriptfont0=\fiverm
\noindent
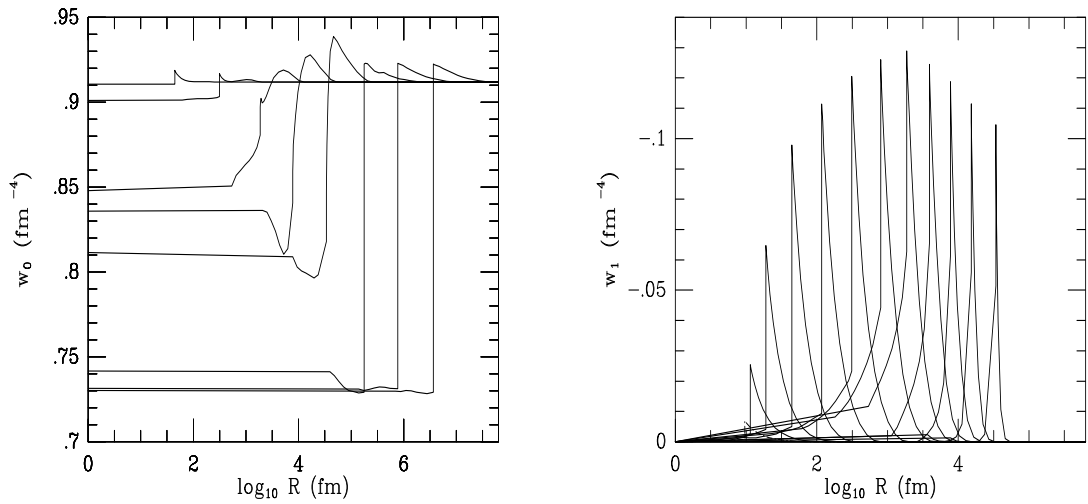
Figures 4--5. Radiation fluid energy density $w_0$ and energy flux
$w_1$. Here $\a_2$ = 1 and $\l$ = 10$^4$ fm. }}

	Next, we turn to a discussion of the effect on the results of
varying the values taken for some of the physical parameters. Bearing in
mind the discussion already given in [7], we will concentrate here just on
the mean-free-path for the radiation particles $\l$ (whose value has an
uncertainty of about an order of magnitude), and on the coefficient $\a_2$
appearing in the expression for the source moment $s_0$ (Eq. (9)).

\vskip 4.0 truecm
\centerline{
\vbox{\hsize=13.0truecm \baselineskip=12pt minus 1pt
\eightrm
\textfont1=\eightmit
\scriptfont0=\fiverm
\noindent
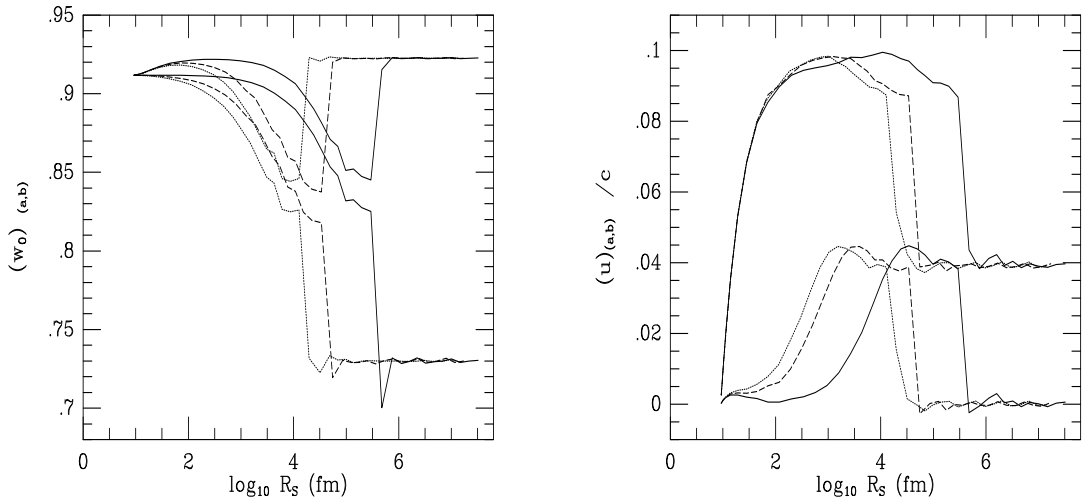
Figures 6--7. Radiation energy density $w_0$ and standard fluid velocity
$u$ just ahead of the phase interface and just behind it (higher and lower
curves respectively). The curves are the result of calculations with
different values of $\l$; $R_{_S}$ is the bubble radius and $\a_2$ = 1.
The curves are plotted so that the hadronic region is on the left of the
discontinuity tracing the phase interface.}} \bigskip

	The effect produced by varying $\l$ is illustrated in Figures 6
and 7, which show the behaviour during the bubble expansion of the values
of $w_0$ and $u$ measured just ahead of the interface and just behind
it. The different values of $\l$ used (which are represented with the
different styles of line) are $5 \times 10^3$ fm, $10^4$ fm and $10^5$
fm, proceeding from left to right. As one would expect, use of a larger
value of $\l$ has the effect of causing the coupling to occur when the
bubble has reached a larger value of $R_s$.

\vskip 4.0 truecm
\centerline{
\vbox{\hsize=13.0truecm \baselineskip=12pt minus 1pt
\eightrm
\textfont1=\eightmit
\scriptfont0=\fiverm
\noindent
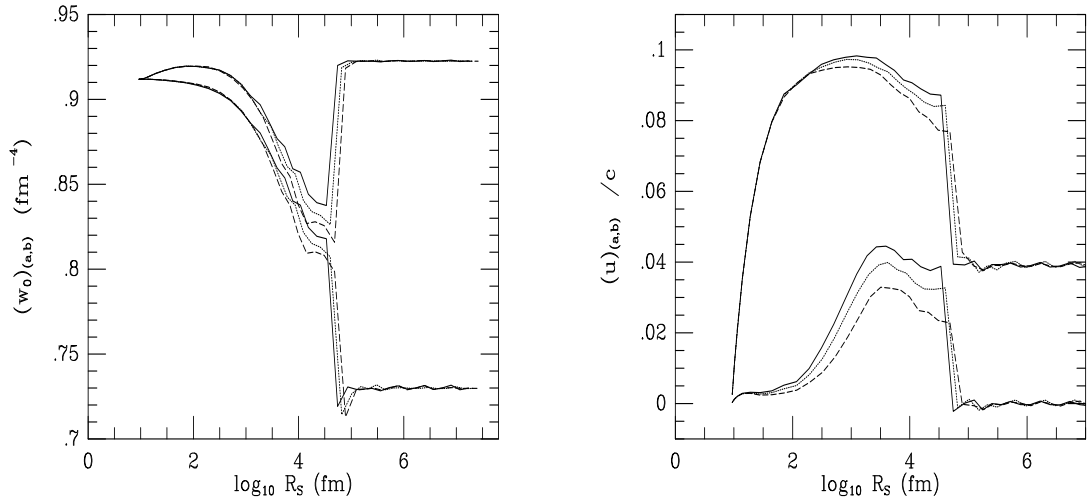
Figures 8--9. Radiation energy density $w_0$ and standard fluid velocity
$u$ just ahead of the phase interface and just behind it. The curves are
the result of calculations with different values of the non-conservative
scattering coefficient $\a_2$. Here \hbox{$\l$ = 10$^4$ fm.} }}
\bigskip

	Figures 8 and 9 show the effect of varying the coefficient for
the non-conservative scatterings ($\a_2$). The different values of
$\a_2$ used are 0, 0.5 and 1; the larger values give an increased rate
of energy transfer and make the total coupling occur earlier.

	These tests show that although the results can indeed be
influenced by different choices of the parameters, there is no serious
qualitative change. We have also checked on the sensitivity of the code to
the form chosen for the Eddington factor $f_{\!_E}$. (The background to
our choice for this was discussed in Section II.) In our investigation, we
have modified the form given in Eq.~(5) by replacing the monotonic
correction term $\l/ (\l+R)$ with a more sophisticated expression having a
maximum whose position and the amplitude could be suitably tuned in
different combinations. Paying attention to producing a smooth join
between the optically-thin and optically-thick limits, we have found that
reasonable variations in this joining function lead to only minor
differences in the results, confirming previous experience [13, 14].
\bigskip
\bigskip
\noindent
{\bf  VI. Conclusion  }
\bigskip
	In this paper, we have presented a study of the hydrodynamics of
the cosmological quark-hadron transition in the presence of long-range
energy and momentum transfer by electromagnetically and weakly
interacting particles. The relativistic radiative transfer problem for
the generalized radiation fluid has been treated using the PSTF tensor
formalism. A system of Lagrangian hydrodynamical equations has been
presented which has then been solved numerically by means of a computer
code which uses a standard Lagrangian finite difference scheme for
flow within the bulk of each phase together with a characteristic method
for the vicinity of the phase interface across which relativistic
junction conditions are solved. The results show the progressive
coupling together of the strongly-interacting matter and the radiation
fluid as the bubble expands. When the complete coupling occurs, there is
no dramatic effect on the bubble which simply decreases its expansion
velocity and the eventually approaches a similarity solution.

	Work is now in progress in order to extend this treatment to the
final stages of the transition during which the evaporation of
disconnected quark droplets occurs. At these stages, which are of great
interest in connection with possible consequences of the transition,
long-range energy and momentum transport certainly play an important
role.
\bigskip
\bigskip
\noindent
{\bf  Acknowledgements}
\bigskip
	We gratefully acknowledge helpful discussions with Ornella
Pantano, Roberto Turolla and Luca Zampieri. Financial support for this
research has been provided by the Italian Ministero dell'Universit\`a
e della Ricerca Scientifica e Tecnologica.
\bigskip
\bigskip
\noindent
\references
\inpress{1} {F. Karsch and E. Laermann} {Rep. Prog. Phys.} {\it in press}
{1994}
\paper{2} {J. Ignatius, K. Kajantie, H. Kurki-Suonio and M. Laine}
{Phys. Rev. D} {49} {3854} {1994}
\paper{3} {J. Ignatius, K. Kajantie, H. Kurki-Suonio and M. Laine}
{Phys. Rev. D} {50} {3738} {1994}
\paper{4} {B. Cheng and A. Olinto} {Phys. Rev. D} {50} {2421} {1994}
\paper{5} {O. Pantano} {Phys. Lett. B} {224} {195} {1989}
\paper{6} {J.C. Miller and O. Pantano} {Phys. Rev. D} {40} {1789} {1989}
\paper{7} {J.C. Miller and O. Pantano} {Phys. Rev. D} {42} {3334} {1990}
\paper{8} {L. Rezzolla and J.C. Miller} {Class. Quantum Grav.} {11} {1815}
{1994}
\paper{9} {H. Kurki-Suonio H} {Phys. Rev. D} {37} {2104} {1988}
\paper{10} {E. Witten} {Phys. Rev. D} {30} {272} {1984}
\paper{11} {K.S. Thorne} {M.N.R.A.S.} {194} {439} {1981}
\paper{12} {K.S. Thorne, R.A. Flammang and A. Zytkow} {M.N.R.A.S.} {194}
{475} {1981}
\paper{13} {L. Nobili, R. Turolla and L. Zampieri} {Ap. J.} {383} {250}
{1991}
\paper{14} {L. Nobili, R. Turolla and L. Zampieri} {Ap. J.} {404} {686}
{1993}
\paper{15} {A. Chodos, R.L. Jaffe, K. Johnson, C.B. Thorn and V.F. Weisskopf}
{Phys. Rev. D} {9} {3471} {1974}
\paper{16} {W. Israel} {Il Nuovo Cimento} {44} {1} {1966}
\paper{17} {K. Maeda} {General Relativity and Gravitation} {18} {931} {1986}

\vfill\eject

\noindent
{\bf Appendix }
\bigskip

	In this Appendix we outline the calculations by means of which the
radiation characteristic equations (41)--(44) are rewritten in the new
form (47)--(49). The aim is to make the ratio ${\ti w_1} / f_{\!_E}$ the
coefficient of a quantity which is small when $f_{\!_E} \to 0$ and in
order to do this we have first isolated the terms in (41) where this ratio
appears:
$$
\displaylines{
{ 4{\ti w_1} \over {3\G f_{\!_E}} }
\Biggl\{ \pm \; \left({1\over 3} +f_{\!_E}\right)^{1/2} du
+ {\Gamma\over {\rho w}}\left(1 -{3\over 2} f_{\!_E} \right) dp \cr
\pm \; a ({1\over 3} +f_{\!_E})^{1/2}
\biggl\{ 4\pi G R\biggl[ p+\biggl({{1\over 3}+
f_{\!_E}}\biggr)w_0 \biggr]+{GM\over {R^2}}+
{\Gamma\over {\rho w}} s_1 \biggr\} dt  \cr
+ a c_s^2 {\Gamma\over {\rho w}}\left(1 -{3\over 2} f_{\!_E} \right)
\biggl[ s_0 + \r w \left( {2u \over R} -
{4 \pi GR w_1 \over {\G}} \right) \biggr] dt \Biggr\}. \cr
}
$$
\rightline{(A1)}
\medskip
\noindent
This can be rearranged to give
$$
\displaylines{
\pm \; { 4 {\ti w_1} \over {3 \G f_{\!_E} }} \Biggl\{
\biggl[ ({1\over 3} +f_{\!_E})^{1/2} - c_s \biggr] du
\mp \; {3 \Gamma\over {2 \rho w}} f_{\!_E}  dp \cr
+ \; a\biggl[ ({1\over 3} +f_{\!_E})^{1/2} - c_s \biggr]
\biggl \{ 4\pi G R\biggl[p+\biggl({{1\over 3}+
f_{\!_E}}\biggr)w_0 \biggr]+{GM\over {R^2}}+
{\Gamma\over {\rho w}} s_1 \biggr \} dt  \cr
\mp \; {3 \Gamma\over {2 \rho w}} f_{\!_E} a c_s^2
\biggl [ s_0 + \r w \left( {2u \over R} -
{4 \pi GR w_1 \over {\G}} \right) \biggr] dt \Biggr\} \cr
\pm \;{4 {\ti w_1} \over {3 \G f_{\!_E} }} c_s \Biggl\{
du\;{\pm}\;{\Gamma\over {\rho w c_s}}dp+
a\biggl\{{\Gamma\over {\rho w}}(s_1\;{\pm}\;c_ss_0) \cr
\hfill{}+4\pi G R\biggl[{p+\biggl({{1\over 3}+
f_{\!_E}}\biggr)w_0\;{\mp}\;c_sw_1}\biggr]+
{GM\over {R^2}}\;{\pm}\;{2\Gamma uc_s\over R}\biggr\} dt \Biggr\}. \cr
}
$$
\rightline{(A2)}
\medskip
\noindent
Replacing (A1) by (A2) then leads to the expressions (47)--(49) which are
the new characteristic equations for the radiation fluid.  Note that the
manipulation has produced two major effects: firstly it has introduced
terms which are intrinsically small when $f_{\!_E} \to 0$ (such as the
difference between the sound speeds or $f_{\!_E} $ itself), and secondly
it has brought together a set of terms which coincides with those on the
left-hand-side of the standard fluid characteristic equation but are now
evaluated along the characteristic directions of the radiation fluid. The
sum of these terms becomes very small when $f_{\!_E}$ is small so that the
two sets of characteristics nearly coincide.
\vfill\eject\end